\documentclass[twocolumn]{svjour3}
  \smartqed
 \usepackage[english]{babel}
 \usepackage[T1]{fontenc}
 \usepackage{times}
 \usepackage[sort,authoryear]{natbib}
 \usepackage{mathptmx}
 \usepackage[dvips]{graphicx}
 \usepackage{amssymb}
 \usepackage{amsmath}				%
 \usepackage{multirow}
 \journalname{Experimental Astronomy}
 \graphicspath{{ART/}}
 \DeclareGraphicsExtensions{.eps,.pdf,.png,.jpg}
\def\smartspace#1{{\protect\aftergroup\smartspaceit#1}}
\def\smartspaceit{\futurelet\spta\sptest}
\def\sptest{\ifcat\noexpand\spta,\else\ \fi}
\let\sspc=\relax
\newcommand{\nc}{\newcommand}

\newcommand{\esm}{\ensuremath}
\newcommand{\mcl}{\multicolumn}
\nc{\etal}{\sspc{et~al.}}
\nc{\eg}{\sspc{e.g.}}
\nc{\etc}{\sspc{etc.}}
\nc{\cf}{\sspc{cf.}}
\nc{\ie}{\sspc{i.e.}}
\nc{\cfig}[1]{\centerline{\psfig{#1}}}
\nc{\Ncol}[1]{\mcl{#1}{c}{~}}
\nc{\mcn}[1]{\mcl{#1}{l}{~}}
\nc{\mcN}[2]{\mcl{#1}{c}{#2}}
\nc{\mcc}[1]{\mcl{1}{c}{#1}}
\nc{\fm}{\esm{\overset{\text{m}}{.}}}
\nc{\dfm}{\esm{\overset{\text{m}}{}}}
 \nc{\AAA}{\sspc{\esm{\lambda\lambda}}}
 \nc{\amm}{\sspc{\,\AA\,mm$^{-1}$}}
 \nc{\kms}{\sspc{\,\esm{\text{km s}^{-1}}}}
 \nc{\msun}{\sspc{\,\esm{\text{M}_{\odot}}}}
 \nc{\rsun}{\sspc{\,\esm{\text{R}_{\odot}}}}
 \nc{\lsun}{\sspc{\,\esm{\text{L}_{\odot}}}}
 \nc{\yr}{\sspc{\,\esm{\text{yr}}}}
 \nc{\kpc}{\sspc{\,\esm{\text{kpc}}}}
 \nc{\halpha}{\sspc{\,\mbox{H$\alpha$}}}
 \nc{\ion}[2]{\sspc{#1\,{\scshape#2}}}
 \nc{\dgr}{\sspc{\esm{^\circ}}}
 \nc{\arcmin}{\sspc{\esm{^{\prime}}}}
 \nc{\arcsec}{\sspc{\raisebox{0.25ex}{\scshape"}}}
 \nc{\Mag}[2]{#1\fm#2}
 \nc{\dMag}[1]{#1\dfm}
 \nc{\dex}[1]{\esm{10^{#1}}}
 \nc{\tdex}[1]{\esm{\times10^{#1}}}
 \nc{\todo}[1]{\sspc{(\textbf{TODO:} #1)}}
 \nc{\obj}[1]{\sspc{#1}}
 \nc{\rotsed}{\sspc{ROTSE--IIId}}
 \nc{\rotse}{\sspc{ROTSE--III}}
 \nc{\tubitak}{\sspc{T\"UB\.ITAK}}
 \nc{\sext}{\sspc{SExtractor}}

\begin{document}
\title{%
  A Pipeline for the \rotsed Archival Data%
}
\titlerunning{%
  Software for \rotsed.%
}
\author{%
B.\,B.\  G\"u\c{c}sav \and
C.\  Ye\c{s}ilyaprak \and
S.\,K.\  Yerli \and
N.\  Aksaker \and
U.\  K{\i}z{\i}lo\u{g}lu \and
D.\  \c{C}oker \and
E.\  Dikicio\u{g}lu \and
M.\,E.\  Ayd{\i}n%
}
\authorrunning{%
  Gucsav B. \etal%
}
\institute{%
  B.\,B.\  G\"u\c{c}sav \and D.\  \c{C}oker \and M.\,E.\  Ayd{\i}n
  \at Ankara \"Universitesi, Science Faculty, Astronomy \& Space Sciences Department, Ankara, Turkey
  \and
  C.\  Ye\c{s}ilyaprak \and E.\  Dikicio\u{g}lu
  \at Atat\"urk University, Faculty of Science, Department of Physics, Erzurum, Turkey
  \and
  S.\,K.\  Yerli \and U.\  K{\i}z{\i}lo\u{g}lu
  \at Orta Do\u{g}u Teknik \"Universitesi, Physics Department, Ankara, Turkey
  \and
  N.\  Aksaker
  \at \c{C}ukurova \"Universitesi, Vocational School of Technical Sciences, Adana, Turkey%
}
\date{%
	Received: date / Accepted: date%
}
\maketitle%

\begin{abstract}
We have constructed a new, fast, robust and reliable pipeline to detect variable stars from the ROTSE-IIId archival data.
Turkish share of ROTSE-III archive contains approximately one million objects from a large field of view (1.85\dgr) and it considerably covers a large portion of northern sky ($\delta>-25\dgr$).
The unfiltered ROTSE-III magnitude of the objects ranges from $7.7$ to $16.9$.
The main stages of the new pipeline are as follows: Source extraction, astrometry of the objects, light curve generation and inhomogeneous ensemble photometry.
A high performance computing (HPC) algorithm has also been implemented into the pipeline where we had a good performance even on a personal computer.
Running the algorithms of the pipeline on a cluster decreases analysis time significantly from weeks to hours.
The pipeline is especially tested against long period variable stars with periods of a few hundred days (e.g Mira and SR) and variables having periods starting from a few days to a few hundred days were detected.
\end{abstract}

\PACS{%
(PACS codes)%
}
\section[%
 Introduction%
]{%
 Introduction%
}
\label{s:10}

Robotic Optical Transient Search Experiment (ROTSE - \citealp{2003PASP..115..132A}) is a network of telescopes located all around the world%
\footnote{\small\url{http://rotse.net/information/world/}}.
The primary goal of the \rotse project is to observe Gamma-Ray Bursts (GRB) in optical light.
Each \rotse telescope consists of 45 cm with a wide field of view (1.85\dgr).
The telescopes were built to respond rapidly to GRBs (<10 s) which are triggered by satellites such as Swift, Integral and HETE.
The \rotse system runs unattended with fully automated observation, data acquisition and analysis (see \cite{2005ApJ...631.1032R} for the pipeline).

The \rotse collaboration uses 70\% of each \rotse telescope's observation time.
The rest of the time is allocated for discretion by the local organization.
The \rotsed telescope is located at \tubitak National Observatory (TUG)%
\footnote{TUG: \small\url{http://www.tug.tubitak.gov.tr/}},
Bak{\i}rl{\i}tepe, Antalya, Turkey.
Scheduled observations were started in May 2004 and they were distributed among the Turkish astronomers by TUG.
In this work, all of the public Turkish observations have been used.
Some of articles which made of using the \rotsed data of Turkish share are as follows \cite{2005A&A...439.1131B}, \cite{2006AN....327..693B} and \cite{2009A&A...508..895K}.

The aim of this work is to detect variability as well as finding new variables using the \rotsed archival data.
To achieve this main goal a multistage algorithm were developed and then they are crafted into a series of routines: \textbf{the pipeline}.
In section \ref{s:rotse}, \rotsed observations and structure of the data, and in section \ref{s:summary} the summary of \rotsed pipeline are given.
The data handling, our pipe\-li\-ne and its structure are given in section \ref{s:30}.
The main stages of the pipeline are as follows: source extraction (\S\ref{s:A}), astrometry of field stars (\S\ref{s:B}), light curve generation (\S\ref{s:C}) and inhomogeneous ensemble photometry (\S\ref{s:D}).
We conclude our work with results and some suggestions for future work.
\section[%
 ROTSE Data%
]{%
 ROTSE Data%
}
\label{s:20}

\subsection{\rotsed observations and structure of the data}
\label{s:rotse}

The \rotsed has an 45 cm primary mirror and is outfitted with a 2k$\times$2k TE cooled, CCD camera with 3.3\arcsec/pixels, making a 1.85\dgr field of view.
QE (Quantum Efficiency) of the CCD peaks at 550 nm.
The telescope and CCD were described in \cite{2003PASP..115..132A}.

The CCD observations have been carried out in three different exposure times: 5 ($\sim 54$\%), 20 and 60 seconds.
There are two important magnitude limits for these exposures given in Table~\ref{T:maglim}: \textbf{saturation} (mean of maximum bright\-ness of stars having no saturated pixels in their Full Width at Half Maximum - FWHM) and \textbf{limiting} (mean of the faintest star's magnitudes).
\begin{table}
\caption{Magnitude limits of \rotsed exposures.}
\label{T:maglim}
\begin{center}
\begin{tabular}{@{}ccc@{}}
Exposure Time &
\mcc{Saturation Magnitude} &
\mcc{Limiting Magnitude} \\ \hline
5  s & \Mag{7}{7} & \Mag{15}{6} \\
20 s & \Mag{9}{0} & \Mag{16}{2} \\
60 s & \Mag{10}{0}& \Mag{16}{9} \\ \hline
\end{tabular}
\end{center}
\end{table}

The unfiltered \rotse magnitude of the objects, depending on the exposure times, ranges from $7.7$ to $16.9$.
Another image quality check was the FWHM of the point spread function on an image which ranges between 2 to 25 pixels and on the average it is taken as 5 pixels.
The Turkish share of \rotsed archive covers 2,210 $\text{deg}^{2}$ which is ~3.4\% of the whole sky.

The journal of observations used in the pipeline ranges between May 2004 and June 2010 and it consists of 234,764 frames from 645 different pointings.
The size of the archived data from the Turkish share (regardless of the pointings) since 2004 is approximately 2 TiB.

266 pointings were observed less than 100 times and they were not included in the pipeline.
This limiting value is an arbitrary choice.
However, to have statistically reliable data sets we had to implement this minimum lower limit.
This number were also used in other large volume \rotse analyses \citep{2004AJ....128.2965W}.

\subsection{The summary of \rotsed pipeline}
\label{s:summary}
\begin{figure}
\centerline{%
\includegraphics[trim=1cm 3cm 1cm 1cm,clip=true,width=\columnwidth]{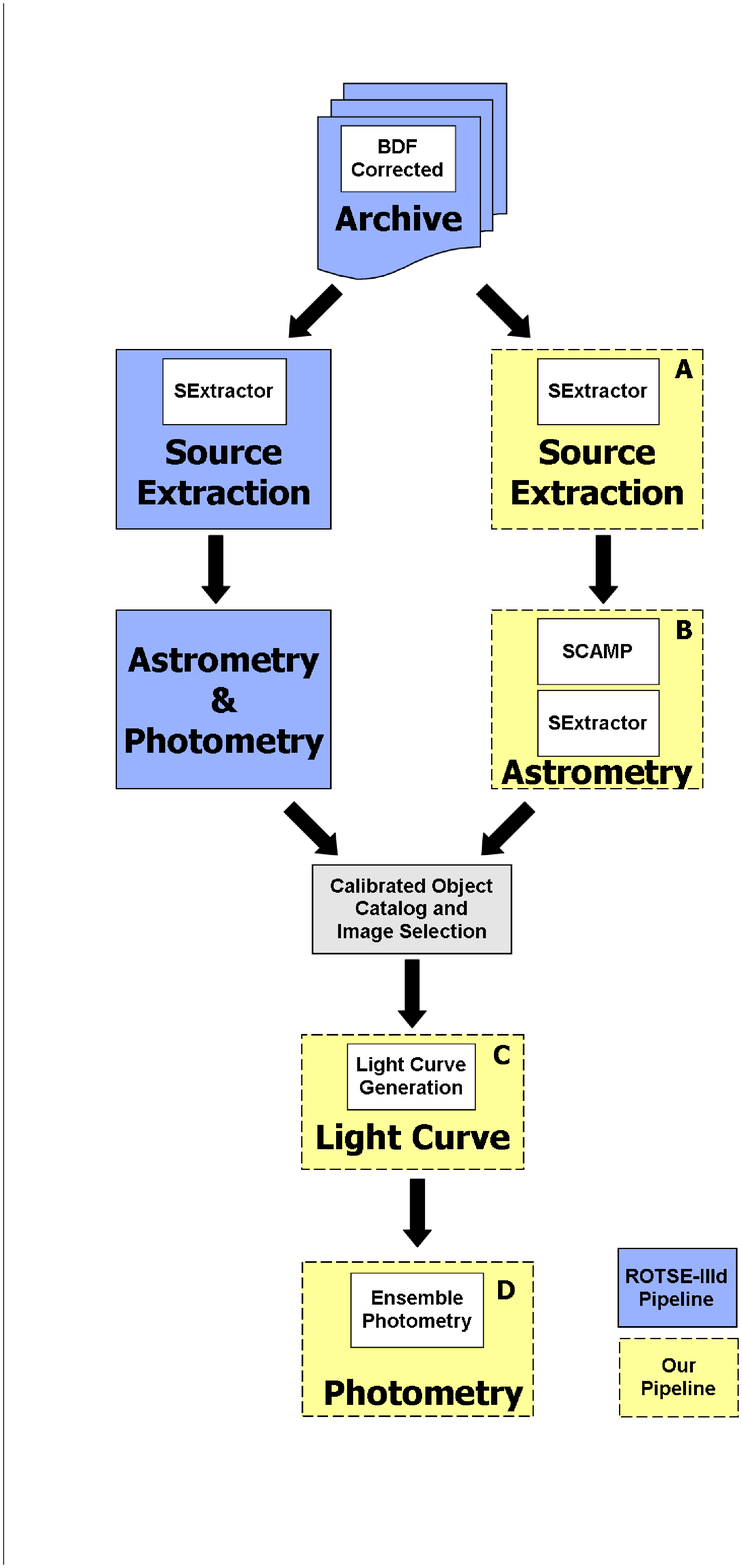}%
}
\caption{%
A combined flow chart view of \rotsed\ (blue colored) and our new (yellow colored) pipelines.%
}
\label{f:Fig1}
\end{figure}

The data acquisition system is constructed on top of several daemons (\eg weather, clamshell, camera).
The telescope is operated under two modes: alarmed and scheduled.
The latter mode is used in the entire Turkish share and thus, in our pipeline.
The \rotse telescopes have a well designed data reduction pipeline allowing the near real-time processing of the CCD images taken by the entire \rotse network.
Regardless of types of observation carried out on the telescope, all images are fed into the \rotsed pipeline.
Furthermore, CCD images of all observations (taken with either alarmed or sched\-uled mode) were automatically processed (bias subtracted, flat fielded and fringe corrected) immediately after the frame has been download to the disk \citep{2005ApJ...631.1032R}.
\sext software \citep{1996A&AS..117..393B} is then used to detect objects, to measure centroid positions and to determine instrumental magnitudes (using 5 pixels aperture).
In addition to centroid positions, roundness and sharpness values are also used to eliminate non-star like objects.
Instrumental magnitudes of each object in the frame is calibrated by comparing \textit{all the field stars} against the ``USNO--A2.0 R--band catalog'' \citep{1998AAS...19312003M} (because the R filter is the nearest to the QE maximum) to obtain \rotse magnitudes.
By simply using triangulation method from the USNO catalog each object's coordinate and instrumental magnitude are calibrated at the same time.
The pipeline finally outputs files of \textbf{calibrated object catalogs} which are tabulated data in the form of binary FITS tables.
The algorithm of the pipeline were coded in IDL and sche\-mat\-ic representation is shown in Figure~\ref{f:Fig1} (blue colored blocks).
Thus, the frame (and each detected object in the frame) can now be used in light curve analysis using object's $\alpha$, $\delta$, magnitude and error in magnitude.
%
%
\section[%
 The New Pipeline%
]{%
 The New Pipeline%
}
\label{s:30}
\subsection{Data Handling}
\label{s:DH}
Input data for the pipeline can either be a calibrated object file or any corrected image file of the \rotsed.
This is required if the frame is not valid \ie the WCS (World Coordinate System) headers were wrong or there were recorded bad weather conditions, or technical problems.
Otherwise the frame is valid and it enters to the stage-C of the pipeline.
Thus, using these filtering methods, at least 80\% of the archive was handled with the pipeline.
The algorithm of the pipeline shown in Figure \ref{f:Fig1} (yellow colored blocks).

It seems that the first two stages of our pipeline duplicates the \rotsed pipeline (see Figure \ref{f:Fig1}).
However, since (a) SCAMP has been used in astrometry calibration (see \S\ref{s:B}) and (b) the parameter set that we use in our pipeline was not produced by the \rotsed, we had to introduce our own stages in the pipeline.

\subsection{Stage-A: Source Extraction}
\label{s:A}

\rotsed corrected images are used as \textbf{INPUT} data in this stage.
A catalog of stars with their instrumental magnitude and frame coordinates is produced as an \textbf{OUTPUT}.

\sext code \citep{1996A&AS..117..393B} is used to identify (and to perform an aperture photometry) the stars in the frame.
Four types of magnitudes are calculated by \sext where only one of them is used in the pipeline, namely \texttt{MAG\_APER} (magnitude from aperture).
A classical value of 5 pixels \rotsed aperture \citep[\eg][]{2005AJ....130.2766K} was used in the \sext settings.
The output of \sext consists of instrumental magnitude, frame coordinates and many statistical calculations for each star.
By using the \sext's statistical output we have improved the reliability of magnitudes calculated which wasn't possible using \rotsed pipeline; it contained only a few columns of information.

As an example; for the \rotsed pointing of 0006+4305 there were 12,037 sources in ``USNO--B1 R'' catalog \citep[hereafter \textbf{USNO--B};][]{2003AJ....125..984M} for $m_R<$\Mag{18}{0}.
When the \sext is applied to this pointing, depending on atmospheric conditions, number of sources varied between 279 and 11,399.
Thus, the stage misses only 5\% of the USNO-B sources.

\subsection{Stage-B: Astrometry of Field Stars}
\label{s:B}

The output of stage-A is used as an \textbf{INPUT}.
Similar to the \rotsed pipeline, the calibrated object catalogs are calculated as an \textbf{OUTPUT}.

The SCAMP \citep{2006ASPC..351..112B} software package is used to map USNO--B stars with the inputted field star coordinates so that a transformation matrix can be calculated.
Instead of USNO--B catalog, GSC, UCAC or 2MASS catalogs could also be used.

SCAMP has been improved for CCD frames taken from large FOV.
The calculated astrometrical error of the \rotsed is approximately 1-2\arcsec for its FOV (less than its pixel size).
Therefore, SCAMP seems to be a suitable tool for astrometric calibrations in our pipeline.

The pointing accuracy (or error) of the \rotsed telescope is given as approximately 1\arcmin \citep{2003PASP..115..132A}.

The mean of differences between ``targeted frame centers'' and ``frame centers recorded in frame headers'' gives the pointing accuracy of \rotsed.
These values are approximately $\overline{\Delta\alpha}=5.1\arcmin$ and $\overline{\Delta\delta}=1.0\arcmin$.
Since unaligned frames within a few arcmin \citep{2006ASPC..351..112B} can be handled with SCAMP, this error doesn't effect the resultant astrometry.

\begin{figure}
\centerline{%
\includegraphics[width=\columnwidth]{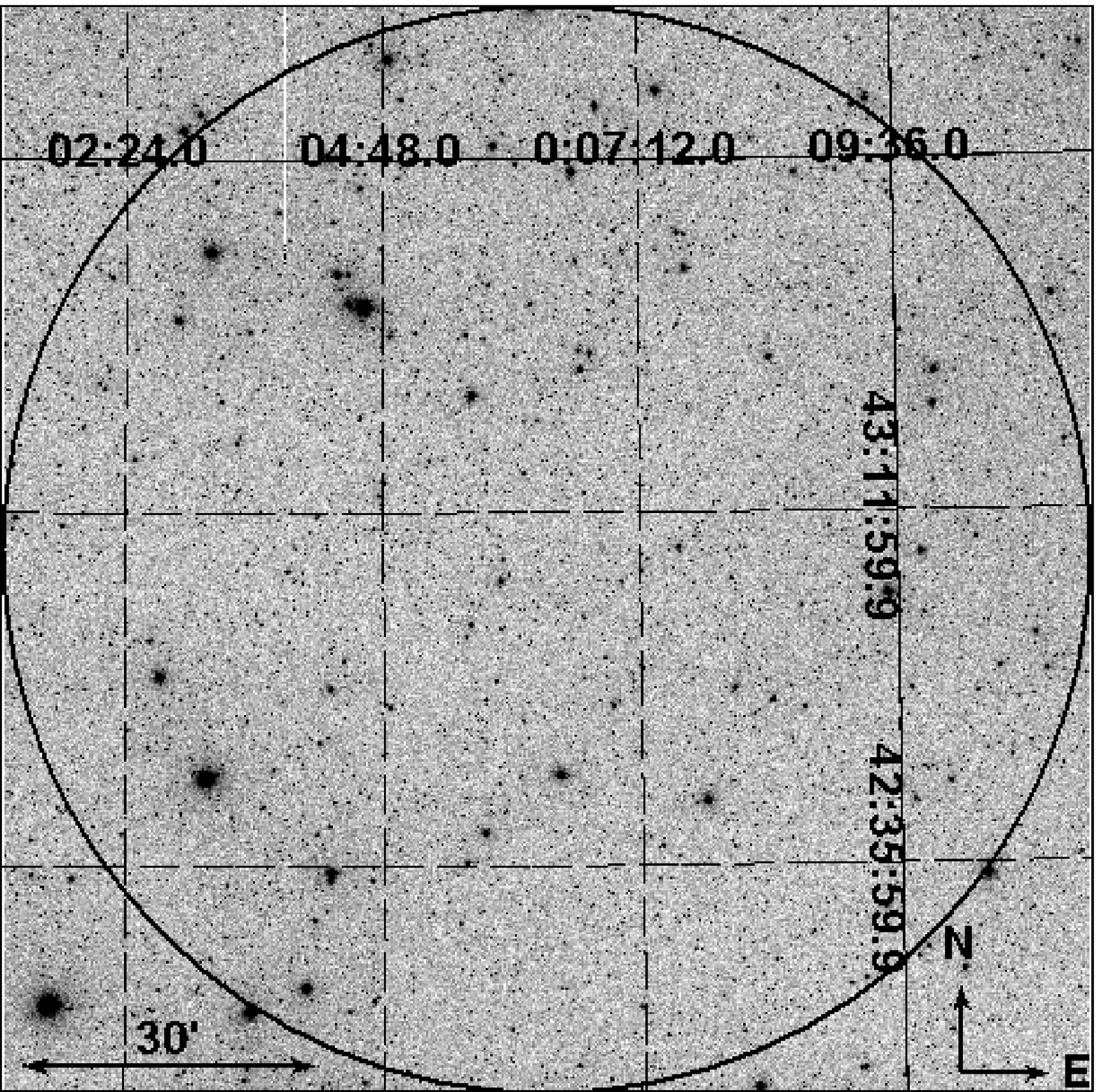}%
}
\caption{%
A sample CCD image from \rotsed archive overlayed with the chosen \rotsed FOV (\ie 1.79\dgr).%
}
\label{f:Fig2}
\end{figure}

Due to the telescope optics \citep{2010MScT.......999G}, the best astrometric alignment of the frames were done within the central 1.79\dgr diameter of the whole \rotsed frame.
Thus, stars falling only into this region were used (see Fig.~\ref{f:Fig2}).
%
%

Since, SCAMP cannot handle frames having excessive image deformations (\eg bad focusing, weather effects, telescope instability) they are automatically discarded and no output is produced.
This feature of SCAMP is also used as a ``filtering'' algorithm for \textit{faulty frames}.

\subsection{Stage-C: Light Curve Generation}
\label{s:C}

The output of stage-B or the output of \rotsed pipeline is used as an \textbf{INPUT}.

Note that, our input contains a list of equatorial coordinates and instrumental magnitudes of sources, namely it is called ``calibrated object catalog'' (see Fig. \ref{f:Fig1}; grey colored box at the middle).
Each inputted star's light curve (instrumental magnitude vs. JD) is created as an \textbf{OUTPUT}.

In this stage, the aim is to follow each star's equatorial coordinates in each frame throughout the whole time span and then collect the corresponding instrumental magnitudes at that coordinate.
As in the stage-B, coordinates of USNO--B field stars are used to match calibrated catalogs.
In order to do this a fixed circular aperture (namely ``matching aperture''; hereafter $\phi_m$) has to be chosen to scan through each catalog.
The $\phi_m$ can be neither a low value (which increases the chance to miss the target star) nor a high value (which increases the chance to multi-match the target stars).
Therefore, $\phi_m$ was chosen according to \rotsed pixel scale (namely 3.3\arcsec/pixel) and to achieve a standard Gaussian photon distribution, 3 pixels range were taken.
Thus, 10\arcsec was chosen as an optimum $\phi_m$ value ($\bar{\phi}_m$) which will be used throughout the pipeline \citep{2010MScT.......999G}.

As a side effect of the matching algorithm, especially in crowded fields, there might still be multi-matches in the field.
In such cases, the nearest star in the calibrated catalog to the USNO--B star was chosen as the match.
The pipeline can be fine-tuned \textit{manually} for some of the overcrowded fields by decreasing $\phi_m$ to 6\arcsec to be able to decrease multi-matches. In a future version, this fine-tuning can be integrated into the pipeline by marking each field with a \textit{crowdness} value which would make it possible to apply different $\phi_m$ values to each field whenever it is necessary.

\subsection{Stage-D: Inhomogeneous Ensemble Photometry}
\label{s:D}

The output of stage-C is used as an \textbf{INPUT}.
\textbf{Cleaned} light curves are created as an \textbf{OUTPUT}.

\begin{figure*}
\centerline{\includegraphics[width=\textwidth]{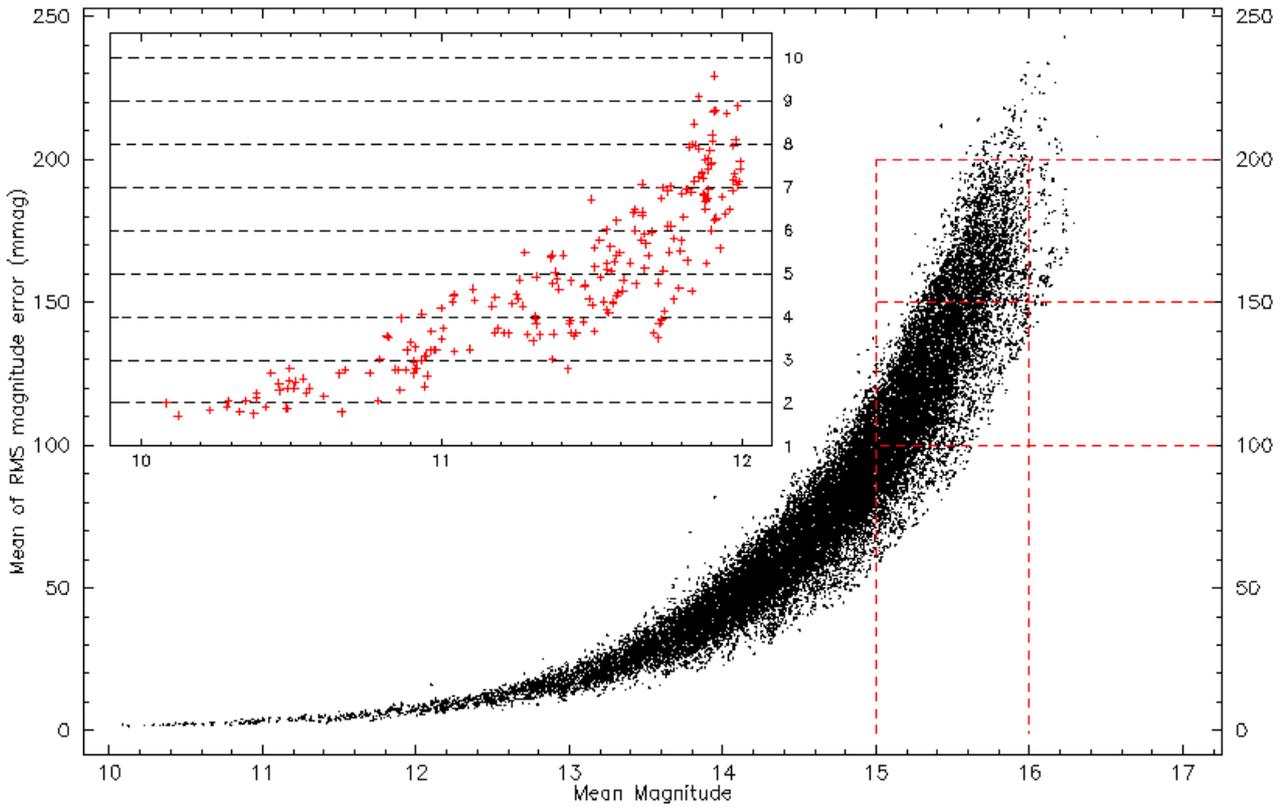}}
\caption{%
Mean magnitude (in magnitude) versus mean of RMS flux errors (in mmag) of sources in an example pointing which contains 31,984 sources is given (see \S\ref{s:D}).
The bright end of the graph is extended at upper left as an inset with the same units.
Similarly, dim end of the graph is marked with grid lines.%
}
\label{f:Fig3}
\end{figure*}

In classical photometry, a single comparison star is used effectively for many years by astronomers \citep[see][]{1990apth.book.....H, 1991PASP..103..221Y}.
However, in recent years differential photometry with ``many comparison stars'' has increased both reliability and accuracy of the light curves.
\textit{Ensemble photometry} is also a new kind of differential photometry which works with inhomogeneous CCD data sets \citep[\eg][]{1992PASP..104..435H, 2010A&A...515A..16S}.
We have also implemented the ensemble photometry in the pipeline to decrease statistical errors in instrumental magnitude of stars.
For example, light curve of a \Mag{10}{0} star was found to be almost constant.
According to \sext, mean of \textit{RMS flux errors} of the PSF fitting to the star was calculated to be \Mag{0}{006}.
However, with the implemented ensemble photometry, we have reached to a scatter value of \Mag{0}{002}.
Note that, this level of noise is also related to the other frame statistics (see scatter-error graph of the pointing that this star is located - Figure \ref{f:Fig3}); as the magnitude of the source decreases, scatter increases and therefore goodness of the PSF fitting decreases.
As can be seen in the figure, while scatter of a \Mag{10}{0} magnitude star is around 2 mmag, it increases to 130--200 mmag at \Mag{16}{0}.
%
%

The main aim of the technique is to find non-varying stars (i.e. reference stars) throughout the time span of the light curves. The objects that have non-star like shapes are all ignored by using \sext's roundness and sharpness analysis (see \S\ref{s:summary}).
Main criteria in choosing the reference stars are (1) to choose the stars far enough from edges of the frame, (2) to have the star isolated from the others, (3) to have no flags set in the \sext's output, (4) to have roundness value close to zero, and (5) to have sharpness value close to one.
In order to not to foul the technique, light curves of mostly non-variable stars have to be used.
Therefore, rough variability detection has to be applied to the light curves; namely `scatter-and-error' relation of the light curve has to be calculated.
According to the result of this relation the star can now be accounted as a reference.
Afterward, a mean reference level is calculated from all chosen reference stars and it is used to calculate the relative (or differential) magnitude of other stars.

\begin{figure}
\centerline{%
\includegraphics[width=\columnwidth]{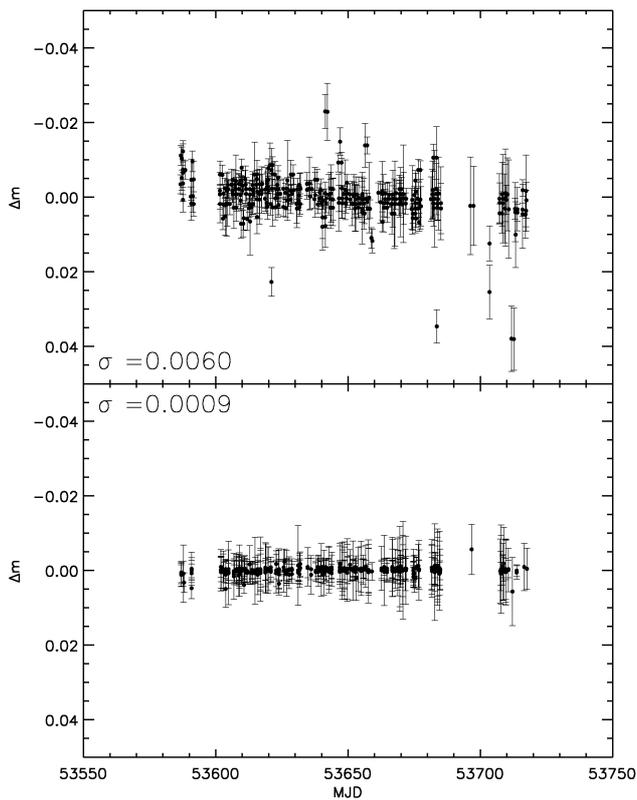}%
}
\caption{%
The light curves of a reference star before (top panel) and after (bottom panel) ensemble photometry is applied.
The sigma values represent the scattering of the curve.%
}
\label{f:Fig4}
\end{figure}

By applying the technique second time (starting from the scatter-and-error relation), reliability of reference stars is increased: stars of the first run is inputted as the star list in the second run.
With this run, variation of reference stars is decreased and therefore the accuracy of the final cleaned light curves is increased.
As an example, a reference star's light curves before and after ensemble photometry applied are given in Fig.~\ref{f:Fig4}.
As seen in the figure, after ensemble photometry (stage-D) applied, scatter of the light curve (\ie sigma) decreased.
%
%

\begin{figure}
\centerline{%
\includegraphics[width=\columnwidth]{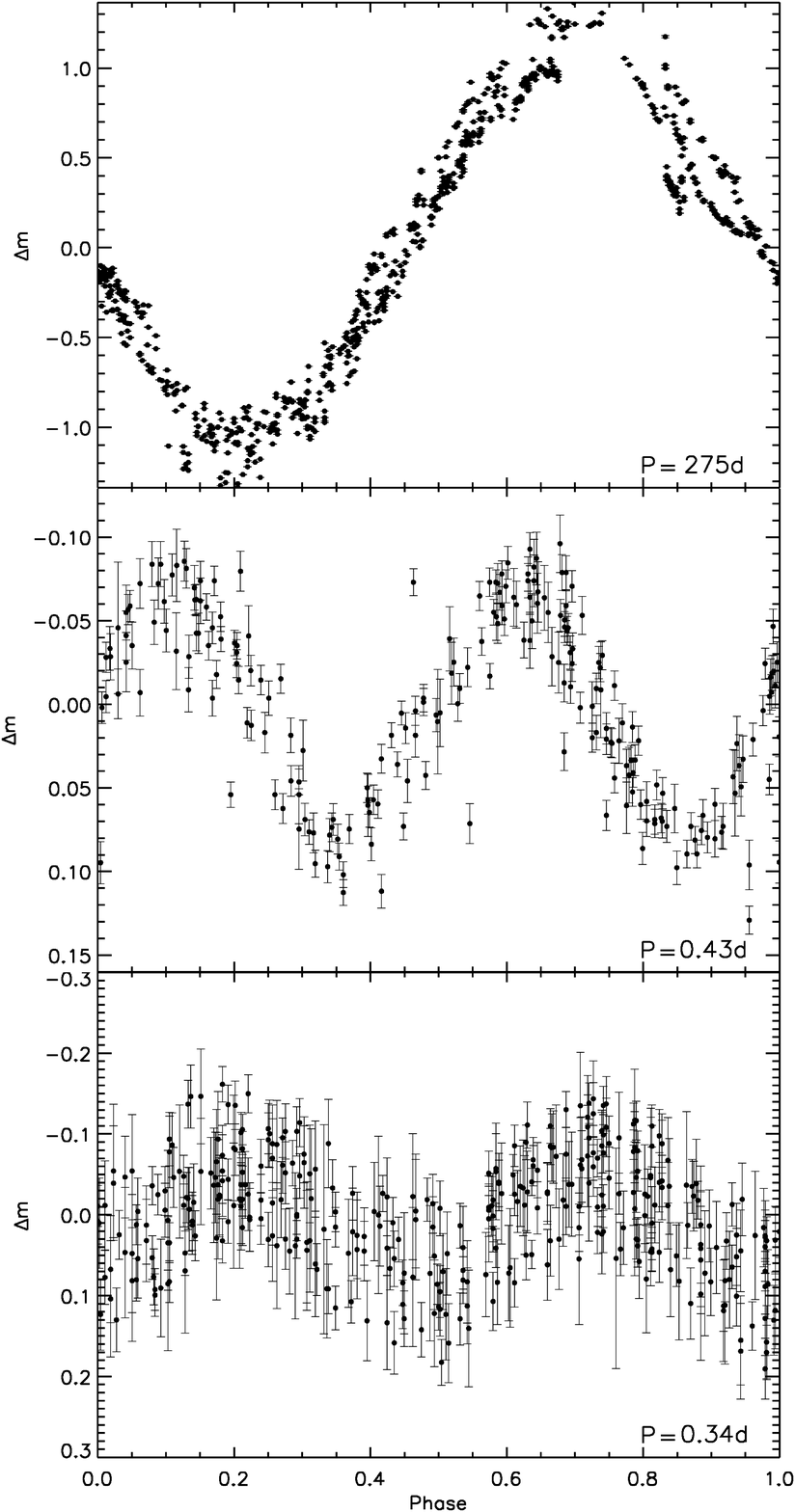}%
}
\caption{%
Light curves (differantial magnitudes vs.\ phase) of three selected variables
which are automaticly produced by our pipeline. They all have different mean
magnitudes and their mean statistical errors are also plotted on each point
(except the top one; which is too small to be plotted together). From top to
bottom panel, the mean magnitudes are \Mag{8}{3}, \Mag{12}{4} and \Mag{15}{5};
and their mean statistical errors are 1 mmag, 10 mmag and 40 mmag,
respectively.%
}
\label{f:Fig5}
\end{figure}

This technique has the advantage of removing frame to frame background variations due to Moon light and unstable weather conditions.
As a disadvantage of this technique, the absolute magnitude of stars cannot be calculated.
A sample of the final light curve of a variable from our pipeline is given in Fig.~\ref{f:Fig5}.
%
%

\subsection{The Pipeline Structure}

The new pipeline works on a small cluster called Infinitus which made use of 8 different computers with 36 cores each having 8 GHz CPU speed.
GNU/Linux operating system and Lustre 1.6.7.4 file system is used on all computers.
Computers are interconnected with a gigabyte ethernet.
The pipeline is mainly written in C language.
Parallel algorithms have been used in every step of the pipeline which made use of MPICH2%
\footnote{\url{http://www.mcs.anl.gov/research/projects/mpich2/}}
library.

In order to increase speed of the process a parallel file reading method is used in the pipeline.
The files of each pointing (all the calibrated catalog stars in the entire time span) is automatically separated into all processors so that each pointing stays in the memory up to end of the stage-C.
In order to balance the CPU load between cores the same amount of star's data (both light curves and star's information) are stored.

The prototype of the pipeline was started with IDL.
However, to be able to have a robust and fast pipeline, they are converted and parallelized into C-language.
With the prototype, it took approximately one week to create light curves of a single medium crowded pointing with a 4 cored PC.
This duration decreased to about 2 hours with the pipeline.
\section[%
 Conclusions%
]{%
 Conclusion%
}
\label{s:40}

We have constructed a new, fast, robust and reliable pipeline to detect variable stars from the \rotsed archival data.
The main stages of the pipeline were as follows: Object identification, astrometry of the objects, light curve generation and inhomogeneous ensemble photometry.
For the first time in the \rotse archive, a high performance computing (HPC) algorithm has been implemented into the pipeline.
Depending on the data quality, either corrected CCD images (stage-A) or the calibrated object catalogs produced by the \rotsed pipeline (stage-C) could be used in the pipeline.
The last stage of the pipeline (stage-D), namely \textit{inhomogeneous ensemble photometry} (implemented to the \rotse archive for the first time), gives relative light variations of the measured stars with high precision.
Within 645 pointings (see section~\ref{s:rotse}) of observations in 2004--2010, light curve of approximately one million stars are produced.

Work on identifying new variables and searching for periods of known variables are still in progress.
The following statistical tests to detect variable stars have been applied to the light curves:
    Scatter-Error Analysis (\textit{commonly used}),
    Abbe Index \citep{2010A&A...515A..16S} and
    Analysis of Variance \citep{2004AJ....128.2965W}.
For period hunting the following techniques have been used:
    PDM (Phase Dispersion Minimization; \citealp{1978ApJ...224..953S}),
    Lomb--Scargle periodogram \citep{1976Ap&SS..39..447L, 1982ApJ...263..835S} and
    SIGnificance SPECtrum (SigSpec; \citealp{2007A&A...467.1353R}).
A few thousands of light curves have passed from all three statistical tests mentioned above.
These light curves most probably are \textbf{un-identified} variable stars.

Light curves of approximately ten thousands stars from 4 pointings are converted into phase-magnitude table.
These phase graphs were then visually inspected for repeating patterns.
Among these reduced list, 152 stars show variability and according to SIMBAD \citep{2010MScT.......999G}, 20 of them are \textbf{unknown} variables which are not classified yet.

As a result of an early version of the pipeline, Mira and SR stars known in SIMBAD have been examined.
New actual periods of 78 Mira variables have been identified and 18 of them
have a period for the first time \citep{2011Mira..........Y} with early
version of the pipeline.
Also, approximately 300 SR stars are under investigation and 15 of them were examined in details by \cite{2011MScT.......999D}.
Tabulated values of minimum and maximum values of both amplitude and period are
given in Table~\ref{T:LClimits}.
\begin{table}
\caption{Quantitative limits of light curves created by the pipeline.}
\label{T:LClimits}
\begin{center}
\begin{tabular}{@{}c@{\hspace*{6pt}}c@{\hspace*{6pt}}c@{\hspace*{6pt}}l@{}}
&
Amplitude &
Period (days) &
Reference \\\hline
Min. & \Mag{0}{1} & 0.17$\pm$0.02 & \cite{2010MScT.......999G}\\
Max. & \Mag{5}{9} &  720.0$\pm$43.1 & \cite{2011Mira..........Y}\\ \hline
\end{tabular}
\end{center}
\end{table}

As a by product of the pipeline, deformation and/or degeneration of frames could also be detected; \eg background variations due to Moon light and unstable weather conditions.
A surprising result from degradation of the \rotsed mirror was also noticed: the decrease in the number of detected stars is correlated with one magnitude decrease in the limiting magnitude.

The main aim of the pipeline usage is to find new variables, periods of known variables, and to classify these variables using the above period hunting techniques.

The pipeline can easily be adapted to other \rotse telescopes which will increase the sky coverage and number of detected unknown variables, transients \etc
\begin{acknowledgements}
This project utilizes data obtained by the Robotic Optical Transient Search Experiment (ROTSE).
ROTSE is a collaboration of Lawrence Livermore National Lab, Los Alamos National
Lab and the University of Michigan (http://www.rotse.net).

All observations were made with the \rotsed telescope and the archival data of
\rotsed obtained at the \tubitak (Turkish Scientific and Research Council)
National Observatory (TUG), so we thank to \rotse Collaboration and TUG for the
optical and archival facilities (TUG - \rotsed projects of Turkish observers).

We also thank Prof. Dr. \"U. K{\i}z{\i}lo\u{g}lu for consulting, suggestions
and helps.

This study was supported by \tubitak with the project TBAG--108T475.

This research has made use of the SIMBAD database, operated at CDS, Strasbourg,
France and cdsclient tool located at CDS and NASA Astrophysics Data System Bibliographic Services.
\end{acknowledgements}


  \label{lastpage}

\begin{thebibliography}{22}
\providecommand{\natexlab}[1]{#1}
\providecommand{\url}[1]{{#1}}
\providecommand{\urlprefix}{URL }
\expandafter\ifx\csname urlstyle\endcsname\relax
  \providecommand{\doi}[1]{DOI~\discretionary{}{}{}#1}\else
  \providecommand{\doi}{DOI~\discretionary{}{}{}\begingroup
  \urlstyle{rm}\Url}\fi
\providecommand{\eprint}[2][]{\url{#2}}

\bibitem[{Akerlof et~al(2003)Akerlof, Kehoe, McKay, Rykoff, Smith, Casperson,
  McGowan, Vestrand, Wozniak, Wren, Ashley, Phillips, Marshall, Epps, and
  Schier}]{2003PASP..115..132A}
Akerlof CW, Kehoe RL, McKay TA, Rykoff ES, Smith DA, Casperson DE, McGowan KE,
  Vestrand WT, Wozniak PR, Wren JA, Ashley MCB, Phillips MA, Marshall SL, Epps
  HW, Schier JA (2003) {The ROTSE-III Robotic Telescope System}. The
  Publications of the Astronomical Society of the Pacific 115:132--140

\bibitem[{{Baykal} et~al(2005){Baykal}, {K{\i}z{\i}lo{\v g}lu},
  {K{\i}z{\i}lo{\v g}lu}, {Balman}, and {Inam}}]{2005A&A...439.1131B}
{Baykal} A, {K{\i}z{\i}lo{\v g}lu} {\"U}, {K{\i}z{\i}lo{\v g}lu} N, {Balman}
  {\c S}, {Inam} S{\c C} (2005) {X-ray outburst of 4U 0115+634 and ROTSE
  observations of its optical counterpart V635 Cas}. Astronomy and Astrophysics
  439:1131--1134

\bibitem[{{Bertin}(2006)}]{2006ASPC..351..112B}
{Bertin} E (2006) {Automatic Astrometric and Photometric Calibration with
  SCAMP}. In: Astronomical Data Analysis Software and Systems XV, Astronomical
  Society of the Pacific Conference Series, vol 351, pp 112--115

\bibitem[{{Bertin} and {Arnouts}(1996)}]{1996A&AS..117..393B}
{Bertin} E, {Arnouts} S (1996) {SExtractor: Software for source extraction.}
  Astronomy and Astrophysics Supplement 117:393--404

\bibitem[{Bilir et~al(2006)Bilir, G{\"u}ver, and Aslan}]{2006AN....327..693B}
Bilir S, G{\"u}ver T, Aslan M (2006) {Separation of dwarf and giant stars with
  ROTSE-IIId}. Astronomische Nachrichten 327:693--697

\bibitem[{{Dikicio\u{g}lu}(2011)}]{2011MScT.......999D}
{Dikicio\u{g}lu} E (2011) {Multiple Periods for Semiregular (SR) Variable
  Stars.} Master's thesis, Atat\"urk University, Erzurum, Turkey

\bibitem[{{G\"u\c{c}sav}(2010)}]{2010MScT.......999G}
{G\"u\c{c}sav} BB (2010) {Detection of Different Types of Celestial Objects
  from Robotic Telescope Archives.} Master's thesis, Ankara University, Ankara,
  Turkey

\bibitem[{{Henden} and {Kaitchuck}(1990)}]{1990apth.book.....H}
{Henden} AA, {Kaitchuck} RH (1990) {Astronomical photometry : a text and
  handbook for the advanced amateur and professional astronomer}

\bibitem[{Honeycutt(1992)}]{1992PASP..104..435H}
Honeycutt RK (1992) {CCD ensemble photometry on an inhomogeneous set of
  exposures}. Astronomical Society of the Pacific 104:435--440

\bibitem[{{K{\i}z{\i}lo{\v g}lu} et~al(1995){K{\i}z{\i}lo{\v g}lu},
  {K{\i}z{\i}lo{\v g}lu}, and {Baykal}}]{2005AJ....130.2766K}
{K{\i}z{\i}lo{\v g}lu} {\"U}, {K{\i}z{\i}lo{\v g}lu} N, {Baykal} A (1995)
  {ROTSE Observations of the Young Cluster IC 348}. The Astronomical Journal
  130:2766--2777

\bibitem[{{K{\i}z{\i}lo{\v g}lu} et~al(2009){K{\i}z{\i}lo{\v g}lu}, Ozbilgen,
  {K{\i}z{\i}lo{\v g}lu}, and Baykal}]{2009A&A...508..895K}
{K{\i}z{\i}lo{\v g}lu} {\"U}, Ozbilgen S, {K{\i}z{\i}lo{\v g}lu} N, Baykal A
  (2009) {Optical and X-ray outbursts of Be/X-ray binary system SAX
  J2103.5+4545}. Astronomy and Astrophysics 508:895--900

\bibitem[{Lomb(1976)}]{1976Ap&SS..39..447L}
Lomb NR (1976) {Least-squares frequency analysis of unequally spaced data}.
  Astrophysics and Space Science 39:447--462

\bibitem[{Monet(1998)}]{1998AAS...19312003M}
Monet DG (1998) {The 526,280,881 Objects In The USNO-A2.0 Catalog}. Bulletin of
  the American Astronomical Society 30:1427

\bibitem[{Monet et~al(2005)Monet, Levine, Canzian, Ables, Bird, Dahn, Guetter,
  Harris, Henden, Leggett, Levison, Luginbuhl, Martini, Monet, Munn, Pier,
  Rhodes, Riepe, Sell, Stone, Vrba, Walker, Westerhout, Brucato, Reid,
  Schoening, Hartley, Read, and Tritton}]{2003AJ....125..984M}
Monet DG, Levine SE, Canzian B, Ables HD, Bird AR, Dahn CC, Guetter HH, Harris
  HC, Henden AA, Leggett SK, Levison HF, Luginbuhl CB, Martini J, Monet AKB,
  Munn JA, Pier JR, Rhodes AR, Riepe B, Sell S, Stone RC, Vrba FJ, Walker RL,
  Westerhout G, Brucato RJ, Reid IN, Schoening W, Hartley M, Read MA, Tritton
  SB (2005) {The USNO-B Catalog}. The Astronomical Journal 125:984--993

\bibitem[{Reegen(2007)}]{2007A&A...467.1353R}
Reegen P (2007) {SigSpec. I. Frequency- and phase-resolved significance in
  Fourier space}. Astronomy and Astrophysics 467:1353--1371

\bibitem[{{Rykoff} et~al(2005){Rykoff}, {Aharonian}, {Akerlof}, {Alatalo},
  {Ashley}, {G{\"u}ver}, {Horns}, {Kehoe}, {Kizilo{\v g}lu}, {McKay},
  {{\"O}zel}, {Phillips}, {Quimby}, {Schaefer}, {Smith}, {Swan}, {Vestrand},
  {Wheeler}, {Wren}, and {Yost}}]{2005ApJ...631.1032R}
{Rykoff} ES, {Aharonian} F, {Akerlof} CW, {Alatalo} K, {Ashley} MCB,
  {G{\"u}ver} T, {Horns} D, {Kehoe} RL, {Kizilo{\v g}lu} {\"U}, {McKay} TA,
  {{\"O}zel} M, {Phillips} A, {Quimby} RM, {Schaefer} BE, {Smith} DA, {Swan}
  HF, {Vestrand} WT, {Wheeler} JC, {Wren} J, {Yost} SA (2005) {A Search for
  Untriggered GRB Afterglows with ROTSE-III}. The Astrophysical Journal
  631:1032--1038

\bibitem[{Saesen et~al(2010)Saesen, Carrier, Pigulski, Aerts, Handler, Narwid,
  Fu, Zhang, Jiang, Vanautgaerden, Kopacki, Steslicki, Acke, Poretti,
  Uytterhoeven, Gielen, Ostensen, De~Meester, Reed, Kolaczkowski, Michalska,
  Schmidt, Yakut, Leitner, Kalomeni, Cherix, Spano, Prins, van Helshoecht,
  Zima, Huygen, Vandenbussche, Lenz, Ladjal, Puga~Antolin, Verhoelst,
  De~Ridder, Niarchos, Liakos, Lorenz, Dehaes, Reyniers, Davignon, Kim, Kim,
  Lee, Lee, Kwon, Broeders, van Winckel, Vanhollebeke, Waelkens, Raskin, Blom,
  Eggen, Degroote, Beck, Puschnig, Schmitzberger, Gelven, Steininger,
  Blommaert, Drummond, Briquet, and Debosscher}]{2010A&A...515A..16S}
Saesen S, Carrier F, Pigulski A, Aerts C, Handler G, Narwid A, Fu JN, Zhang C,
  Jiang XJ, Vanautgaerden J, Kopacki G, Steslicki M, Acke B, Poretti E,
  Uytterhoeven K, Gielen C, Ostensen R, De~Meester W, Reed MD, Kolaczkowski Z,
  Michalska G, Schmidt E, Yakut K, Leitner A, Kalomeni B, Cherix M, Spano M,
  Prins S, van Helshoecht V, Zima W, Huygen R, Vandenbussche B, Lenz P, Ladjal
  D, Puga~Antolin E, Verhoelst T, De~Ridder J, Niarchos P, Liakos A, Lorenz D,
  Dehaes S, Reyniers M, Davignon G, Kim SL, Kim DH, Lee YJ, Lee CU, Kwon JH,
  Broeders E, van Winckel H, Vanhollebeke E, Waelkens C, Raskin G, Blom Y,
  Eggen JR, Degroote P, Beck P, Puschnig J, Schmitzberger L, Gelven GA,
  Steininger B, Blommaert J, Drummond R, Briquet M, Debosscher J (2010)
  {Photometric multi-site campaign on the open cluster NGC 884. I. Detection of
  the variable stars}. Astronomy and Astrophysics 515:A16

\bibitem[{Scargle(1982)}]{1982ApJ...263..835S}
Scargle JD (1982) {Studies in astronomical time series analysis. II -
  Statistical aspects of spectral analysis of unevenly spaced data}.
  Astrophysical Journal 263:835--853

\bibitem[{{Stellingwerf}(1978)}]{1978ApJ...224..953S}
{Stellingwerf} RF (1978) {Period determination using phase dispersion
  minimization}. Astrophysical Journal 224:953--960

\bibitem[{Wozniak(2004)}]{2004AJ....128.2965W}
Wozniak WSJVWTGV P~R (2004) {Identifying Red Variables in the Northern Sky
  Variability Survey}. The Astronomical Journal 128:2965--2976

\bibitem[{{Ye\c{s}ilyaprak} et~al(2011){Ye\c{s}ilyaprak}, {Yerli},
  {G\"u\c{c}sav}, {Aksaker}, {Dikicio\u{g}lu}, {Helvac{\i}}, {\c{C}oker},
  {Ayd{\i}n}, {Din\c{c}el}, and {Uzun}}]{2011Mira..........Y}
{Ye\c{s}ilyaprak} C, {Yerli} S, {G\"u\c{c}sav} B, {Aksaker} N, {Dikicio\u{g}lu}
  E, {Helvac{\i}} M, {\c{C}oker} D, {Ayd{\i}n} M, {Din\c{c}el} B, {Uzun} N
  (2011) {Long-Term Variations and Periods of Mira Stars from \rotsed.}
  Accepted by New Astronomy

\bibitem[{{Young} et~al(1991){Young}, {Genet}, {Boyd}, {Borucki}, {Lockwood},
  {Henry}, {Hall}, {Smith}, {Baliumas}, {Donahue}, and
  {Epand}}]{1991PASP..103..221Y}
{Young} AT, {Genet} RM, {Boyd} LJ, {Borucki} WJ, {Lockwood} GW, {Henry} GW,
  {Hall} DS, {Smith} DP, {Baliumas} SL, {Donahue} R, {Epand} DH (1991) {Precise
  automatic differential stellar photometry}. Astronomical Society of the
  Pacific 103:221--242

\end{thebibliography}
\end{document}